\documentclass{aa}

\usepackage{graphicx}
\usepackage{txfonts}
\usepackage[citecolor=blue, linkcolor=blue, urlcolor = black, colorlinks = true]{hyperref}

\usepackage{graphicx}	
\usepackage{amsmath}	
\usepackage{amssymb}	
\usepackage{xspace}

\usepackage{siunitx}
\usepackage{comment}
\usepackage{textcomp}
\usepackage{bm}
\usepackage{lscape}
\usepackage{xcolor}

\newcommand{\gdor}{$\gamma$~Dor\xspace}

\newcommand{\Msun}{\,M$_{\odot}$\xspace}

\newcommand{\xc}{$X_{\rm c}$\xspace}
\newcommand{\fov}{$f_{\rm CBM}$\xspace}

\newcommand{\percent}{~per~cent\xspace}

\newcommand{\omegai}{$(\Omega_{\rm surf}/\Omega_{\rm crit})_{\rm ZAMS}$\xspace}

\newcommand{\ca}[1]{#1}

\newcommand{\pin}{$\Pi_0$\xspace}
\newcommand{\edit}[1]{#1}

\newcommand{\mesa}{\texttt{MESA}\xspace}

\newcommand{\ester}{\texttt{ESTER}\xspace}
\begin{document}

   \title{Probability distributions of initial rotation velocities and core-boundary mixing efficiencies of $\gamma$~Doradus stars  }

   \author{J.S.G. Mombarg\inst{1}
   \and 
   C. Aerts\inst{2,3,4}
   \and 
   G. Molenberghs\inst{5,6}
          }
   \institute{IRAP, Universit\'e de Toulouse, CNRS, UPS, CNES, 14 avenue \'Edouard Belin, F-31400 Toulouse, France\\
   \email{jmombarg@irap.omp.eu}
   \and 
   Institute of Astronomy, KU Leuven, Celestijnenlaan 200D, 3001 Leuven, Belgium
   \and
   Department of Astrophysics, IMAPP, Radboud University Nijmegen, PO Box 9010, 6500 GL, Nijmegen, The Netherlands
   \and
   Max Planck Institute for Astronomy, Koenigstuhl 17, 69117 Heidelberg, Germany
   \and
   I-BioStat, Universiteit Hasselt, Martelarenlaan 42, 3500 Hasselt, Belgium
   \and
   I-BioStat, KU Leuven, Kapucijnenvoer 7, 3000 Leuven, Belgium              
        }

   \date{Received 11 January 2024; accepted 7 February 2024}
\titlerunning{Statistical modelling of \gdor stars}
\authorrunning{Mombarg et al.}
 
  \abstract
   {The theory the rotational and chemical evolution is incomplete, thereby limiting the accuracy of model-dependent stellar mass and age determinations. The $\gamma$~Doradus (\gdor) pulsators are excellent points of calibration for the current state-of-the-art stellar evolution models, as their gravity modes probe the physical conditions in the deep stellar interior. Yet, individual asteroseismic modelling of these stars is not always possible because of insufficient observed oscillation modes. }
  { This paper presents a novel method to derive distributions of the stellar mass, age, core-boundary mixing efficiency and initial rotation rates for \gdor stars. }
   {We compute a grid of rotating stellar evolution models covering the entire \gdor instability strip. We then use the observed distributions of the luminosity, effective temperature, buoyancy travel time and near-core rotation frequency of a sample of 539 stars to assign a statistical weight to each of our models. This weight is a measure of how likely the combination of a specific model is. We then compute weighted histograms to derive the most likely distributions of the fundamental stellar properties. }
   {We find that the rotation frequency at zero-age main sequence follows a normal distribution, peaking around 25\percent of the critical Keplerian rotation frequency. The probability-density function for extent of the core-boundary mixing zone, given by a factor \fov times the local pressure scale height (assuming an exponentially decaying parameterisation) decreases linearly with increasing \fov.}
   {Converting the distribution of fractions of critical rotation at the zero-age main sequence to units of d$^{-1}$, we find most F-type stars start the main sequence with a rotation frequency between 0.5\,d$^{-1}$ and 2\,d$^{-1}$. Regarding the core-boundary mixing efficiency, we find that it is generally weak in this mass regime.}


   \maketitle
%
\section{Introduction}
The state-of-the-art stellar structure and evolution models that lay at the basis of many stellar-ageing methods lack a complete picture of chemical mixing. In case of stars that maintain a convective core during the main sequence (MS) phase, our ignorance regarding the efficiency of chemical mixing is often parameterised by a function for the core-boundary mixing \citep[CBM; e.g.][]{Zahn1991,Freytag1996, Augustson2019} and a function for the mixing in the radiative envelope \citep[e.g.][]{Pedersen2021}, both with at least one free parameter. The exact choice for these free parameters can result in age differences of about 40\percent by the end of the main sequence \citep{Mombarg_thesis}, or up 15\percent between models with and without time-dependent self-consistent convective penetration \citep{Johnston2023}. In addition, chemical mixing is induced by rotational shear \citep[e.g.][]{Zahn1992} and therefore we require an accurate description for the transport of angular momentum. Yet, confrontations of predictions of angular momentum transport with asteroseismic measurements of the rotation velocities of stars have shown that the current physics is not adequate \citep{Eggenberger2012, Marques2013, cantiello2014, Aerts2019-ARAA}. 

The class of $\gamma$~Doradus (\gdor) gravity- (g) mode pulsators \citep{kaye1999} has proven to be useful to constrain the near-core rotation \citep[][]{VanReeth2016, christophe2018, Li2019, Li2020}, the (radial) differential rotation \citep{VanReeth2018, ouazzani2020, Saio2021}, and stellar mass and age \citep[e.g.][]{Mombarg2019, Mombarg2021, Mombarg2023}. As such, these constraints can be used to test the theory of angular momentum \citep{Ouazzani2019, Moyano2023}. In particular, \cite{Mombarg2023} has tested a diffusive approach for angular momentum transport on a set of slow rotating \gdor pulsators by combining their measured rotation frequencies from \cite{Li2019} with asteroseismic masses, ages and CBM efficiencies using the method of \cite{Mombarg2021}. When testing angular momentum transport, assumptions have to be made about the initial rotation. In the study of \cite{Ouazzani2019}, the rotation frequency at the zero-age main sequence (ZAMS) is estimated from a stellar disk model with free parameters calibrated to cluster data, while in the studies of \cite{Moyano2023} and \cite{Mombarg2023} the (uniform) rotation at the ZAMS is left as a free parameter. \cite{Mombarg2023} concludes that the six slowly-rotating stars in his sample also had a slow rotation (less than 10\percent of the critical rotation frequency) to begin with. However, the distribution of rotation frequencies at the ZAMS for \gdor pulsators or F-type stars in general is not well known, and this paper aims to improve on that. 

Similarly, the distribution of the efficiency of the CBM that needs to be added to obtain the core masses inferred by asteroseismic modelling is not well known \citep{Johnston2021, Pedersen2022a}. Individual measurements of the CBM efficiency have been made, by modelling the observed mode frequencies of main-sequence g-mode pulsators \citep[e.g.][]{johnston2019, Mombarg2021, Michielsen2019, Szewczuk2022, Michielsen2023}, and of subgiants \cite[e.g.][]{Deheuvels2011,Deheuvels2016, Noll2021}. The first step (after frequency extraction) in the modelling of a g-mode pulsator is identifying the spherical degree ($\ell$), azimuthal order ($m$) and radial order ($n$) of the excited pulsation frequencies. This is done by exploiting the relation that consecutive radial orders with the same ($\ell, m$)-combination form a pattern when the difference in period between consecutive modes as plotted against the mode period itself \citep{Tassoul1980, miglio2008, bouabid2013}. Once such a period-spacing pattern is identified, the near-core rotation frequency and buoyancy travel time (\pin, an asteroseismic quantity related to the g-mode cavity) can be measured, as first put into practice by \cite{degroote2010}. 

The current largest sample of \gdor pulsators with identified period-spacing patterns is that of \cite{Li2020}, comprising 611 stars. In the case of single \gdor stars, a precise measurement of the CBM efficiency requires both a sufficient number of identified radial orders and a precise constraint on the effective temperature. Apart from a subsample of 37 stars with spectroscopically derived effective temperatures already modelled by \cite{Mombarg2021}, a large part of the sample of \cite{Li2020} does not allow for precise constraints on the CBM efficiency. However, as also shown by \cite{Garcia2022b}, measurements of the near-core rotation frequency and buoyancy travel time (\pin) can be made more easily. 

The aim of this paper is to present a novel method to place constraints on the CBM efficiency and initial rotation velocity using the complete sample of \gdor pulsators. This methodology is based on modelling the distributions of the observed luminosity, effective temperature, buoyancy travel time, and near-core rotation frequency of \gdor pulsators. Relying on distributions instead of individual measurements makes this method less susceptible to individual uncertainties, as long as the sample size is sufficiently large.

\section{Statistical methodology} \label{sec:method}
In this paper, we make use of the largest sample of \gdor to-date from \cite{Li2020}, comprising 611 stars observed with the NASA {\it Kepler} mission \citep{borucki2010}. We are interested in finding the distributions of the stellar mass, $M_\star$, the hydrogen-mass fraction in the core (proxy for age), \xc, the efficiency of CBM, \fov, (further discussed in Sections~\ref{sec:models} and \ref{sec:distributions}), and the rotation velocity at zero-age main sequence (ZAMS) as a fraction of the Keplerian critical rotation frequency, 
\ca{$\omega_0\equiv$ \omegai.} As observables, we have the distributions of the luminosity, $L_\star$, derived from the Gaia DR2 parallax \citep{Murphy2019}, of the effective temperature, $T_{\rm eff}$, from \citep{Mathur2017}, of the buoyancy travel time, \pin, and of the near-core rotation frequency probed by the g~modes, $\Omega_{\rm core}$.
The buoyancy travel time is defined as,
\begin{equation}
    \Pi_0 = 2 \pi^2 \left( \int_{\rm gc} N\,{\rm d}\ln r \right)^{-1},
\end{equation}
and the near-core rotation frequency as,
\begin{equation}
    \Omega_{\rm core} = \frac{\int_{\rm gc} \Omega N\,{\rm d}\ln r }{\int_{\rm gc} N \,{\rm d}\ln r}.
\end{equation}
Here, $N$ is the Brunt-V\"ais\"al\"a frequency, $\Omega$ the local angular rotation velocity, and $r$ the radial coordinate. Both integrals are evaluated over the g-mode cavity, defined as the region where the mode frequency in the corotating frame is smaller than $N$. Spectroscopically-derived effective temperatures are available for only about 50 of the 611 stars \cite{Gebruers2021}. Therefore, we rely on the photometric ones. 

\begin{figure*}
    \centering
    \includegraphics[width = \textwidth]{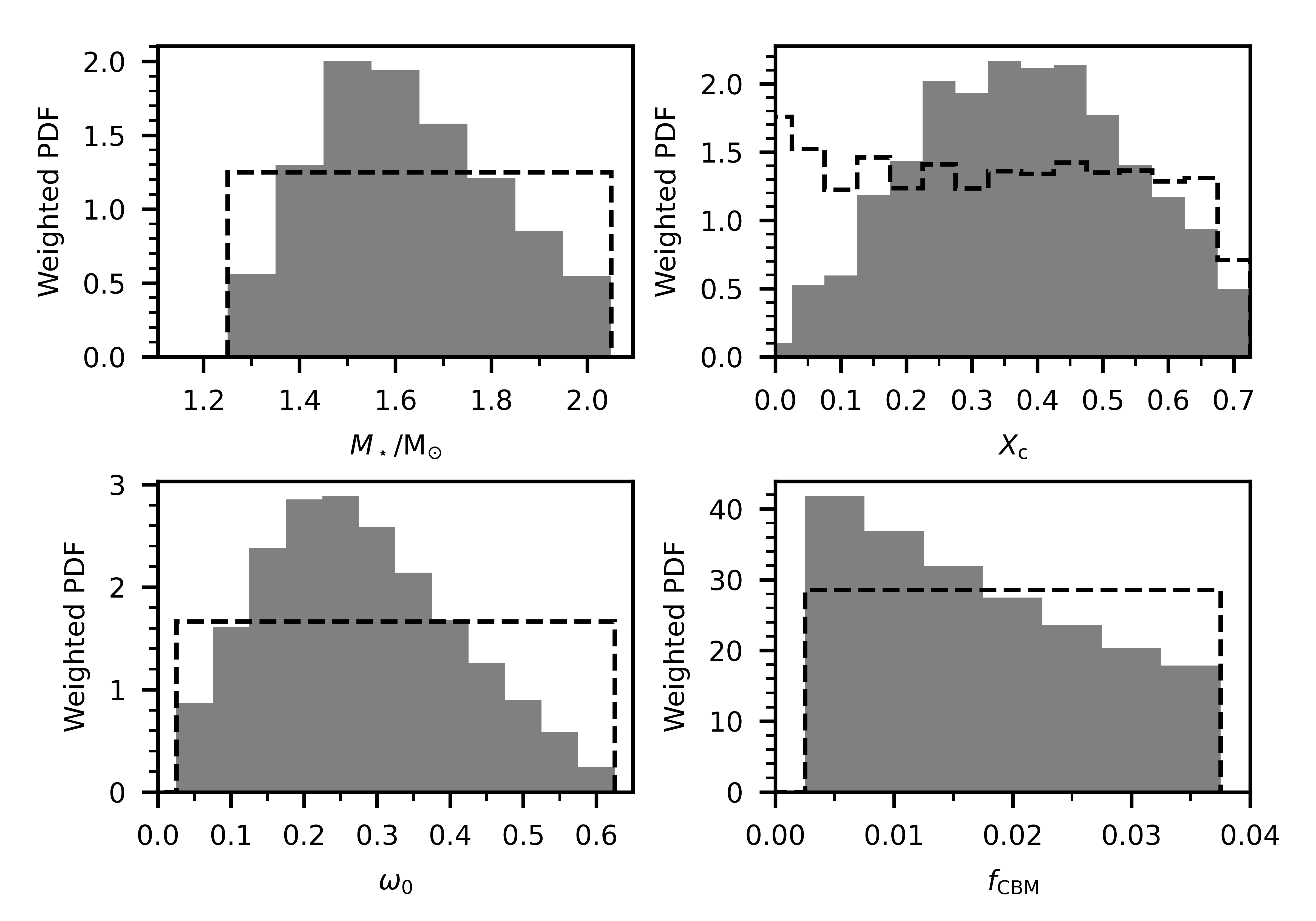}
    \caption{Weighted probability density functions of the stellar mass (top-left), the hydrogen-mass fraction in the core (top-right), the fraction of critical rotation at the ZAMS (bottom-left), and the CBM efficiency (bottom-right). The black dashed lines show the occurrence of each value in the grid ($\hat{x}_{\rm m}$). }
    \label{fig:distr}
\end{figure*}

\edit{To estimate the most likely values of the four fundamental parameters,}
given the distributions of the observed quantities and their precisions, we compute weighted histograms, for which we define a weight $\tilde{\rho}$ as follows. First, let $\hat{y}_{{\rm o}, i} = (L_{\star, i}, T_{{\rm eff}, i}, \Pi_{0,i}, \Omega_{{\rm core}, i})$ be a vector containing the observed quantities of star $i \in [1, ..., N]$. We then compute a mean vector $\hat{\mu}$ and \ca{variance-covariance} matrix $\hat{\Sigma}$,
\begin{eqnarray}
\hat{\mu} &=& \frac{1}{N}\sum_i^n \hat{y}_{{\rm o}, i}, \\
\hat{\Sigma} &=& \frac{1}{n-1}\sum_i^{n} (\hat{y}_{{\rm o}, i} - \hat{\mu})(\hat{y}_{{\rm o}, i} - \hat{\mu})^{\top}.
\end{eqnarray}
The weight of a model with observables $\hat{y}_{\rm m}$ is then given by,
\begin{equation} \label{eq:rho}
    \tilde{\rho}(\hat{y}_{\rm m}) = \frac{1}{(2\pi)^2 |\hat{\Sigma}|^{1/2}} \exp \left(-\frac{1}{2} (\hat{y}_{\rm m} - \hat{\mu})^{\top} \hat{\Sigma}^{-1} (\hat{y}_{\rm m} - \hat{\mu}) \right)
\end{equation}
\ca{
\citep[see][for a basic introduction into multivariate data analysis]{Johnson2002-stat}.
}
For each model $\hat{x}_{\rm m} = (M_\star, X_{\rm c}, f_{\rm CBM}, \omega_0)$ producing observables $\hat{y}_{\rm m}$, the count towards a bin is weighted by $\tilde{\rho}(\hat{y}_{\rm m})$. We normalise each of the components of $\hat{y}_{\rm m}$ and $\hat{y}_{{\rm o}}$ by the corresponding maximum value in the observed distribution to ensure each of the four observables contributes equally to the value of $\tilde{\rho}$. Furthermore, we take the same number of \xc variations per model, such that the distributions of all components of $\hat{x}_{\rm m}$ are uniformly sampled (see black dashed lines in Fig.~\ref{fig:distr}).  \newline

The sample of \cite{Li2020} contains stars for which there is no luminosity available, or have a measured of \pin larger than 6000\,s and are thus more likely to be Slowly-Pulsating B-type (SPB) stars \citep{Waelkens1991,Pedersen2021}. Furthermore, any known binary system is also excluded. This constitutes 31 stars based on absence of luminosity measurement, 31 based on a too high value of \pin, and 10 stars based on binarity. This leaves us with a final sample of 539 stars. 

\section{Stellar models} \label{sec:models}
In order to compute the vectors $\hat{y}_{\rm m}$ for a given set of model parameters $\hat{x}_{\rm m}$, a grid of stellar models was computed with \mesa \citep[r23.05.1;][]{Paxton2011, Paxton2013, Paxton2015, Paxton2018, Paxton2019, Jermyn2023}\footnote{Our \mesa setup can be found at \url{https://zenodo.org/records/10629035}}. The three fundamental parameters that are varied are the mass, the surface rotation velocity at the ZAMS as a fraction of the critical rotation frequency, and the core-boundary mixing efficiency. For the latter, an exponentially decaying parameterisation of the chemical diffusion parameter is chosen, following \cite{Freytag1996},
\begin{equation} \label{eq:fov}
    D_{\rm CBM}(r) = D(r_0)\exp\left( \frac{-2(r-r_0)}{f_{\rm CBM}h_P(r_{\rm core})}\right).
\end{equation}
Here, $h_P(r_{\rm core})$ is the pressure scale height at the radius of the convective core and $r_0$ is set to $r_{\rm core} - 0.005h_P(r_{\rm core})$. The parameter $f_{\rm CBM}$ is a free parameter determining the efficiency of the CBM. \ca{One of} the aims of this paper is to find a distribution for this parameter, along with the mass, age, and initial rotation velocity. 

Table~\ref{tab:grid} shows the range and step size for each of the four parameters in the grid,
\ca{which contains $\sim$10\,000 models (i.e. different combinations for $\hat{x}_{\rm m}$). }
The chemical mixing in the radiative envelope is based on the work of \cite{Zahn1992}, using the \mesa implementation of \cite{Mombarg2022}. The chemical diffusion coefficient is determined by,
\begin{equation}
    D_{\rm rot}(r>r_{\rm core}) = \eta\ K\ \left( \frac{r}{N} \frac{{\rm d}\Omega}{{\rm d}r}\right)^2, 
\end{equation}
where $K$ is the thermal diffusivity \ca{and $\eta$ is a free parameter. As actual local chemical diffusion coefficient, we take} the largest one out of $D_{\rm CBM}(r), D_{\rm rot}$, or the one from convection. We set the parameter $\eta$ to 1. As discussed in \cite{Mombarg2022}, the shear profile 
\ca{${\rm d}\Omega/{\rm d}r$} 
is scaled from a 2D \ester model \citep{EspinosaLara2013, Rieutord2016} that computes the differential rotation in a self-consistent manner. This way, a smooth profile of the Brunt-V\"ais\"al\"a frequency is ensured, which is necessary for computing asteroseismic quantities. We also include microscopic diffusion (gravitational settling and radiative levitation) for all elements with available monochromatic opacities from the OP Project \citep{Seaton2005}, using the method outlined in \cite{Mombarg2022} and \cite{Jermyn2023}. 
\ca{In doing so}, we neglect the feedback of the change in the local mixture due to microscopic diffusion in the computation of the Rosseland mean opacity, which is \ca{calculated from} the tables of the OP Project.   

We assume a fixed solar metallicity of 0.014 for our models, as spectroscopic studies of samples of {\it Kepler} \gdor stars  have shown that the average metallicity is close to 
\ca{the solar value} \citep[e.g.][]{VanReeth2015b, Kahraman2016, Gebruers2021}. 
\ca{For the initial chemical composition of the models, $(X_{\rm ini}, Y_{\rm ini}, Z_{\rm ini})$, we} assume a
\ca{galactic chemical} enrichment rate $Y_{\rm ini} = 0.244 + 1.226Z_{\rm ini}$ \citep{Verma2019}, giving $Y_{\rm ini} = 0.261$ and $X_{\rm ini} = 0.725$. For the relative metal fractions, we assume the solar mixture according to \cite{Asplund2009}. 

For each model, we compute a non-rotating pre-MS model and relax this model to the desired rotation velocity at the ZAMS. The efficiency of angular momentum transport (given by the viscosity) is computed within the diffusive approach of \mesa \citep[see][for the physical descriptions]{Heger2000}, which includes dynamical shear instability, secular shear instability, Eddington-Sweet, Solberg-Høiland instability, Goldreich-Schubert-Fricke instability and Spruit-Tayler dynamo.  

\begin{table}[htb]
    \centering
    \caption{Ranges and step sizes of the different parameters varied in the \mesa grid.}
    \begin{tabular}{llll}
        \hline \hline
        Parameter & Lower limit & Upper limit & Step size \\
        \hline
        $M_\star/{\rm M_\odot}$ & 1.3 & 2.0 & 0.1 \\
    \ca{$\omega_0$} & 0.05 & 0.60 & 0.05 \\
        $f_{\rm CBM}$ & 0.005 & 0.035 & 0.005 \\
        $X_{\rm c}$ & 0.00 & \edit{0.70} & 0.05 \\
        \hline
    \end{tabular}
    \label{tab:grid}
\end{table}

\section{Inferred distributions} \label{sec:distributions}
For each model in the grid
we compute a weight $\rho(\hat{y}_{\rm m})$ according to Eq.~(\ref{eq:rho}). Then, for $M_\star$, \xc, \fov, and $\omega_0$ we plot the weighted probability density distributions (PDFs). The top-left panel of Fig.~\ref{fig:distr} shows the resulting distribution of the stellar mass, which is \edit{a skewed distribution} around a mass of 1.5-1.6\Msun. It should be noted that the \gdor phenomenon only occurs during a part of a star's main-sequence lifetime. Based on theoretical predictions for the \gdor instability strip, stars with masses close to the edges of the grid (around 1.2 or 2\Msun) will spend a smaller fraction of their main-sequence lifetime within the instability strip compared to stars around 1.6\Msun. Therefore, we indeed expect such a \edit{skewed distribution} distribution instead of recovering the initial-mass function. Yet, the fact that a non-negligible fraction of stars have a mass $>1.8$\Msun suggests that the blue edge of the theoretically-predicted \gdor instability strip \citep{dupret2005} should in reality be extended to higher effective temperatures. \newline

\begin{figure}
    \centering
    \includegraphics[width = \columnwidth]{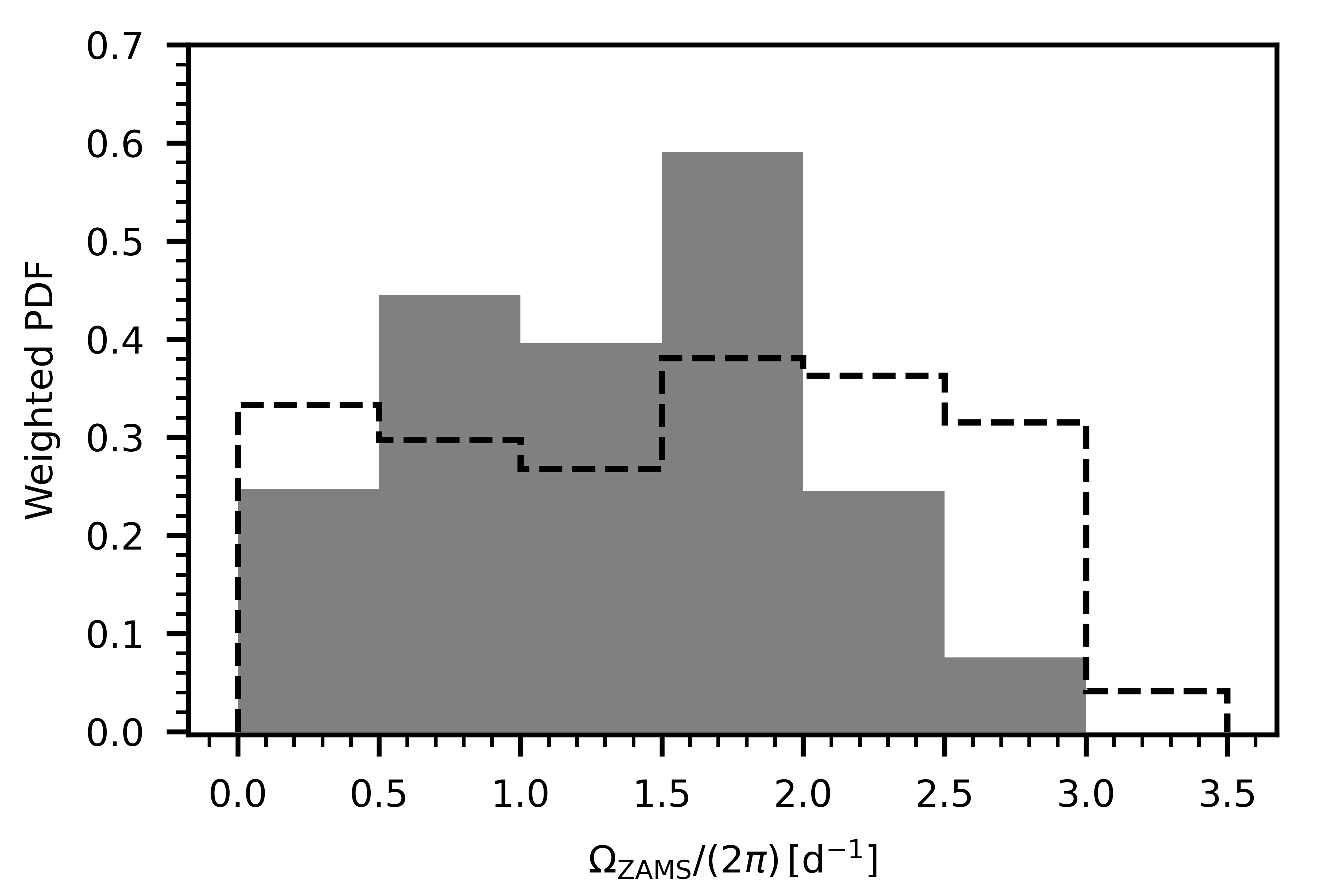}
    \caption{Probability density function of the uniform rotation rate at the ZAMS in units of ${\rm d^{-1}}$. The black dashed line shows the distribution of the models in the grid, the grey histogram the weighted distribution. }
    \label{fig:distr_omega_cpd}
\end{figure}

The bottom-left panel of Fig.~\ref{fig:distr} shows the distribution of the rotation frequencies at the ZAMS as a fraction of the Keplerian critical rotation frequency (at ZAMS). We recover a \edit{skewed distribution} of the PDF that is centred around $\omega_0 = 0.25$. The study by \cite{Li2020} shows an excess of stars with very low near-core rotation frequencies ($<0.15\,{\rm d^{-1}}$) for which \cite{Mombarg2023} concludes these stars were born as slow rotators ($\omega_0<0.1$). From the distribution of $\omega_0$ we recover here, no excess of slow rotators is observed. We also studied the distribution of the (uniform) rotation frequency at ZAMS when we do not scale it with the critical rotation frequency. The distribution is shown in Fig.~\ref{fig:distr_omega_cpd}. As can been seen from the dashed black line in this figure, a uniform distribution of the mass and $\omega_0$, does not give a uniform distribution in $\Omega_{\rm ZAMS}$.

\begin{figure}
    \centering
    \includegraphics[width = \columnwidth]{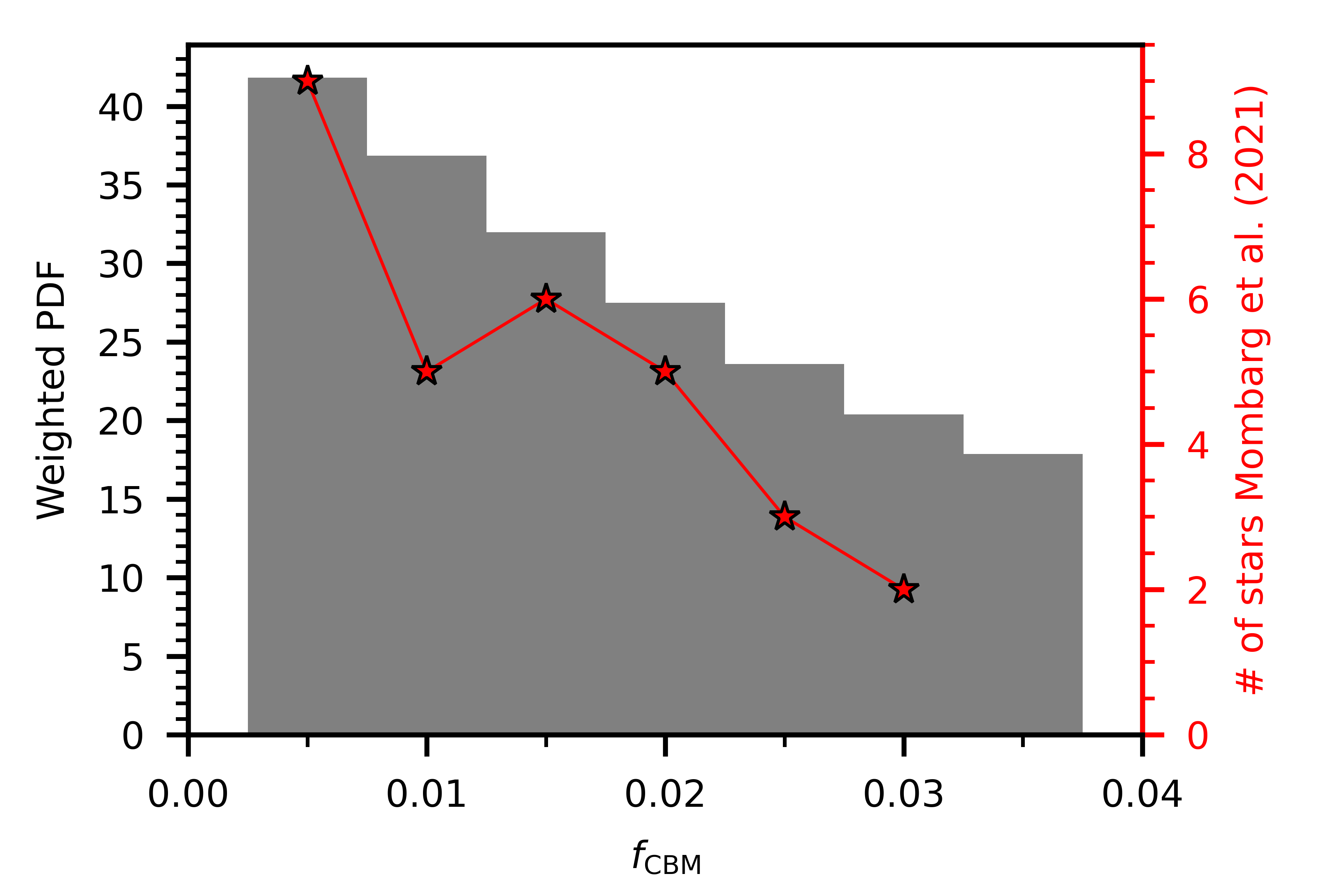}
    \caption{Same probability density function as the bottom-right panel of Fig.~\ref{fig:distr} overplotted with the results of the individually measured \fov values from \cite{Mombarg2021} (red stars).}
    \label{fig:distr_fov_compare}
\end{figure}

\ca{Irrespective of the choice for the binning of the weighted PDF for $\Omega_{\rm ZAMS}$, we conclude that most \gdor stars reach the ZAMS with a rotation frequency between 0.5 and 2$\,{\rm d^{-1}}$ (5.8-23.2$\,{\mu{\rm Hz}}$), although tails at lower and higher rotation frequency are populated as well.} 
\cite{Mombarg2021} estimated the distribution of the initial rotation at the ZAMS of a sample of 37 \gdor stars. Combining the present-day measured near-core rotation frequency \citep{VanReeth2016} with the asteroseismic mass and age, an estimate for the rotation at ZAMS can be made, assuming uniform rotation throughout the MS. When looking at the distribution of the initial rotation frequencies found by \cite{Mombarg2021} a peak around $2\,{\rm d^{-1}}$ is also seen. 
We find no stars with $\Omega_{\rm ZAMS}/(2 \pi) \gtrsim 3\,{\rm d^{-1}}$. This corresponds to the upper limit of the measured near-core rotation frequencies of the \gdor stars from \cite{Li2020} that were used in this paper (see also Fig.\,6 in the summary plot by \cite{Aerts2021}).

\ca{\citet{Fritzewski2024} performed modelling of the solar-metallicity young open cluster UBC\,1 from TESS and Gaia space data and found an age between 150 and 300 Myr. This cluster is much younger than the {\it Kepler\/} field $\gamma\,$Dor stars we used to deduce the distributions. UBC\,1 includes one $\gamma\,$Dor member having a near-core rotation frequency measurement, with a value of 0.544$\pm$0.009\,d$^{-1}$, in agreement with our results for the distribution of rotation rates near the ZAMS. On the other hand, measurements of the near-core rotation frequency of \gdor stars in the even younger open cluster NGC~2516 ($102~\pm~15$~Myr, also solar metallicity) reveal 8 of the 11 g-mode pulsators to have near-core rotation frequencies around 3\,d$^{-1}$, while the 3 others have values between 1\,d$^{-1}$ and about 2.2\,d$^{-1}$ \citep{Li2023-cluster}. At this estimated cluster age, these stars should be even closer to the ZAMS than the $\gamma\,$Dor member of UBC\,1. The rotation rates for the majority of them occurs in the tail of our distribution in Fig.~\ref{fig:distr_omega_cpd}.
Therefore, very young cluster \gdor stars seem to occupy the full range of ZAMS rotation frequencies covered by the distribution we derived from
the much older {\it Kepler\/} field stars in the galaxy and may have a less peaked distribution than the one in Fig.~\ref{fig:distr_omega_cpd} given the possible selection bias for the youngest $\gamma\,$Dor pulsators in the \citet{Li2020} sample, as further discussed in Fritzewski et al.\ (in prep.)}. Finally, with the physics of angular momentum transport used in this paper, we observe models reaching critical rotation during the MS, when $\omega_0 \gtrsim 0.5$.  \newline

The bottom-right panel of Fig.~\ref{fig:distr} shows the distribution of the CBM parameter, \fov (Eq.~(\ref{eq:fov})). We observe a maximum probability density at the lower edge of the grid, $f_{\rm CBM} = 0.005$, and a probability density that decreases linearly with \fov. We find that a value of 0.005 is about twice as likely as a value of 0.035. Thus, larger values are less likely, yet not negligible from a statistical point-of-view. Therefore, we can conclude that a universal value for \fov is not reality, as also advocated by \cite{Johnston2021}. The study of \cite{Mombarg2021} presents individual measurements of \fov (same physical prescription as we use here, apart from the prescription for the envelope mixing) for a sample of 37 \gdor stars. For 30 stars in their sample, the value of \fov could be constrained within the ranges of their grid. We show their distribution of \fov in Fig.~\ref{fig:distr_fov_compare}. Interestingly, the distribution of \cite{Mombarg2021} seems to also follow a linear decrease with the value of \fov. 

\section{Influence of input parameters} \label{sec:parameters}
\begin{figure*}
    \centering
    \includegraphics[width = \textwidth]{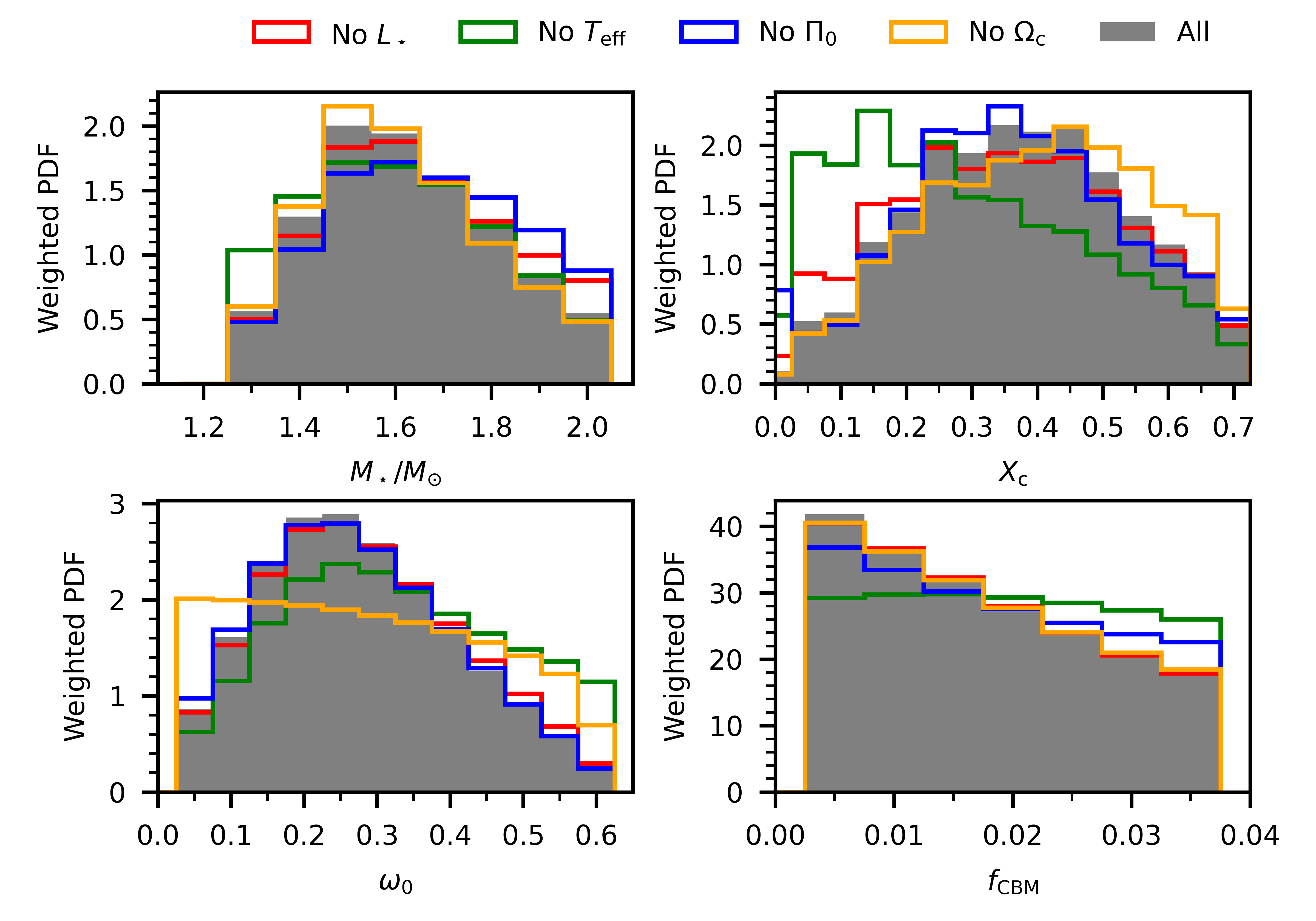}
    \caption{Same as Fig.~\ref{fig:distr}, but each time one of the observables in $\hat{y}_{\rm o}$ is not used to obtain the probability distributions. The histograms in grey are the same ones as shown in Fig.~\ref{fig:distr}.  }
    \label{fig:distr_3p}
\end{figure*}

In this section, we quantify the contribution of each of the four observables in $\hat{y}_{\rm m}$ to the final PDFs. As such, we repeat the methodology of Section~\ref{sec:method}, but omitting one of the observables at a time. From the resulting PDFs shown in Fig.~\ref{fig:distr_3p} we can draw the following conclusions. Firstly, it is obvious that the distribution of $\omega_0$ is mostly determined by the present-day distribution of the near-core rotation frequencies, and that the other fundamental stellar parameters are only mildly influenced. This is expected as there is no feedback of the rotation on the envelope mixing in our models. Secondly, omitting the luminosity results in a slightly higher PDF for more massive stars which are expected to show \gdor pulsations at the end of the MS, thus also increasing the probability for more evolved stars. Thirdly, the value of \pin is sensitive to the core mass \citep[e.g.][]{Mombarg2019} and thus omitting this observable has a large impact on the PDF of \fov. Moreover, since the value of \pin drops off rapidly near the TAMS, including this observable eliminates the peak in stars near the TAMS, as shown in the top-right panel of Fig.~\ref{fig:distr_3p}. Finally, we see that the effective temperature is an important parameter to break degeneracies of \pin with respect to mass and age. \edit{The effective temperature has an even larger effect on the resulting PDF of \fov compared to \pin. Increasing the extent of the CBM zone results in a higher effective temperature at the same luminosity. Therefore, without an effective temperature, models with a larger value for \fov become equally likely, resulting in a flatter PDF. }

\section{Conclusions} \label{sec:conclusions}
In this paper, we have presented a novel method to obtain stellar mass, age, core-boundary mixing efficiency and initial rotation frequency distributions of pulsating F-type (\gdor) stars. We have used the observed distributions of the luminosity, effective temperature, buoyancy travel time and near-core rotation frequency of a sample of 539 stars 
\ca{with high-precision estimates of these four variables taken from the {\it Kepler\/} asteroseismic $\gamma\,$Dor catalogue published by \citet{Li2020}.} We computed a grid of rotating (1D) stellar models for different masses, ages, core-boundary mixing efficiencies and initial rotation velocities at the ZAMS and assigned a statistical weight to each stellar model to obtain the probability density functions of these four fundamental stellar parameters. This method allows us to also include stars that are not suitable for \ca{asteroseismic modelling of the individual mode frequencies. The }  method is robust against individual measurement errors, as long as the sample is large enough to accurately sample average values and covariances of the observables. 

The distributions presented in this paper can be used as priors for future  modelling using a Bayensian framework 
or for population synthesis of pulsating F-type stars. We find \edit{skewed distributions} of the probability density function for the mass, hydrogen-mass fraction in the core (\xc, proxy for age), and fraction of critical rotation at the ZAMS ($\omega_0$). These distributions peak around 1.6\Msun, $X_{\rm c} = 0.4$, and $\omega_0 = 0.25$. We find the probability distribution of the extent of the core-boundary mixing region (assuming an exponentially decaying function) to be linearly decreasing with increasing \fov. The results on the initial rotation and core-boundary mixing presented in this paper are consistent with results from star-by-star modelling of the individual observed mode periods performed by \cite{Mombarg2021}.  

The method presented in this paper could also be applied to 
the SPB class of gravity mode pulsators. Currently, the sample size of SPB stars with measured \pin values and near-core rotation \ca{frequencies} is about a factor \ca{ten} smaller \citep[see][and references therein]{Pedersen2022b} compared to the \gdor stars. Fortunately, this number is expected to increase \ca{from more extensive data sets being assembled by the NASA TESS mission \citep{Ricker2015} and the upcoming ESA PLATO mission \citep{rauer2014}.    } 

\newpage
\begin{acknowledgements}
  The research leading to these results has received funding from the French Agence Nationale de la Recherche (ANR), under grant MASSIF (ANR-21-CE31-0018-02). The computational resources and services used in this work were provided by the VSC (Flemish Supercomputer Center), funded by the Research Foundation - Flanders (FWO) and the Flemish Government department EWI. 
   CA acknowledges financial support from the KU\,Leuven Research Council (grant C16/18/005: PARADISE) and from         the European Research
    Council (ERC) under the Horizon Europe programme (Synergy Grant
    agreement N$^\circ$101071505: 4D-STAR). While partially funded by the European Union, views and opinions expressed are however those of the author(s) only and do not necessarily reflect those of the European Union or the European Research Council. Neither the European Union nor the granting authority can be held responsible for them.  
    \edit{The authors are grateful to the anonymous referee for their feedback.}
  This research made use of the \texttt{numpy} \citep{Harris2020} and \texttt{matplotlib} \citep{Hunter2007} \texttt{Python} software packages.   
\end{acknowledgements}

%
%

\bibliographystyle{aa} 
\bibliography{main} 

\appendix

\end{document}